# A 240 Elements Matrix Probe with Aberration Mask for 4D Carotid Artery Computational Ultrasound Imaging

Yuyang Hu, *Student member, IEEE*, Michael Brown, *Member, IEEE*, Didem Dogan, *Member, IEEE*, Mahé Bulot, Maxime Cheppe, Guillaume Ferin, *Member, IEEE*, Geert Leus, *Fellow, IEEE*, Antonius F.W. van der Steen, *Fellow, IEEE*, Pieter Kruizinga, Johannes G. Bosch, *Member, IEEE*

***Abstract*—** **Three-dimensional (3D) ultrasound provides enhanced visualization of the carotid artery (CA) anatomy and volumetric flow, offering improved accuracy for cardiovascular diagnosis and monitoring. However, fully populated matrix transducers with large apertures are complex and costly to implement. Computational ultrasound imaging (cUSi) offers a promising alternative by enabling simplified hardware design through model-based reconstruction and spatial field encoding. In this work, we present a 3D cUSi system tailored for CA imaging, consisting of a 240-element matrix probe with a 40 × 24 mm² large aperture and a spatial encoding mask. We describe the system's design, characterization, and image reconstruction. Phantom experiments show that computational reconstruction using matched filtering (MF) significantly improves volumetric image quality over delay-and-sum (DAS), with spatial encoding enhancing lateral resolution at the cost of reduced contrast ratio. LSQR-based reconstruction was demonstrated to further improve resolution and suppress artifacts. Using both Hadamard and 16-angle plane wave transmission schemes, the system achieved high-resolution images with reasonable contrast, supporting the feasibility of 4D CA imaging applications.**

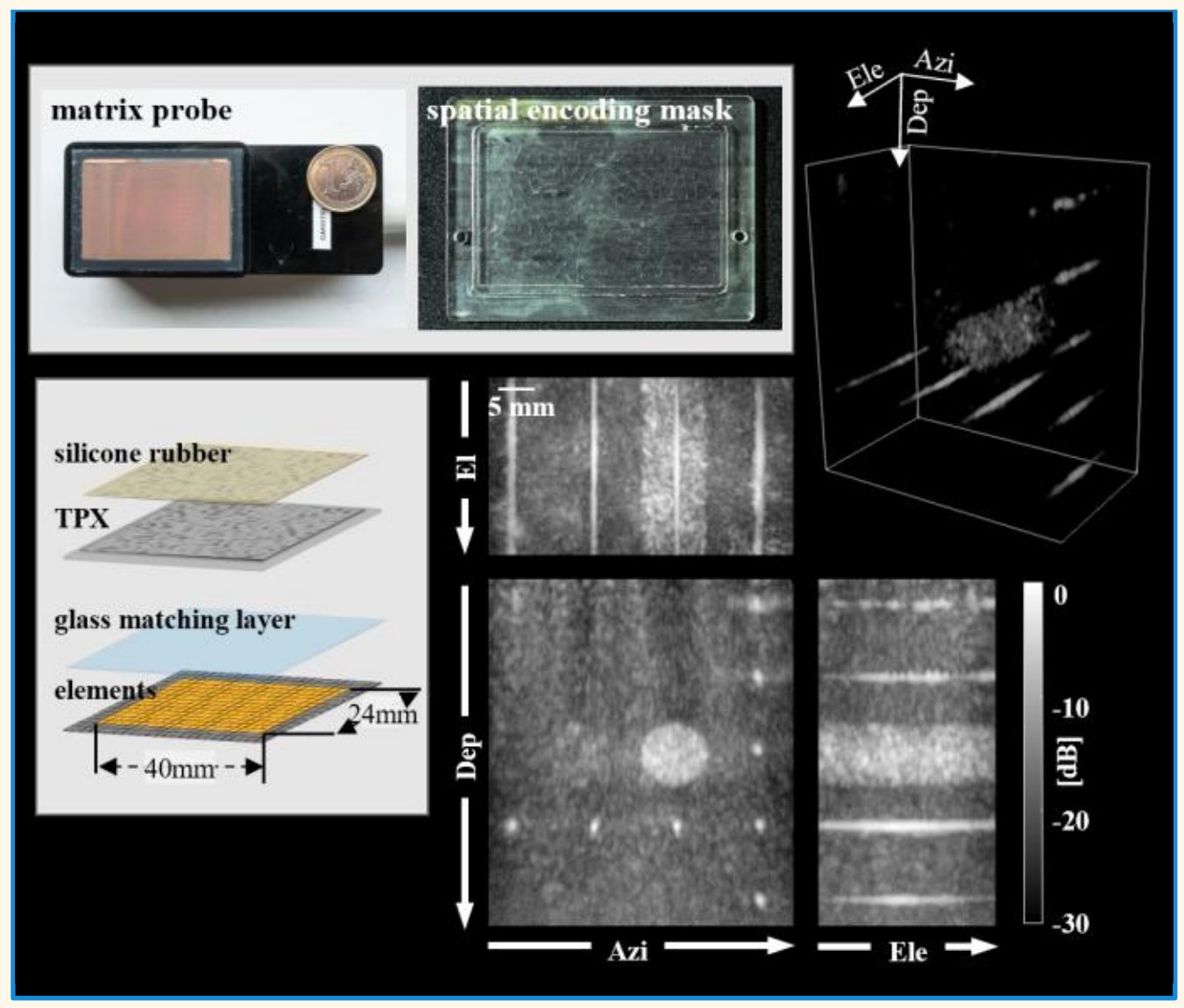

## I. Introduction

THE carotid artery (CA) is highly accessible for noninvasive ultrasound examination and is valuable for cardiovascular health diagnostics or monitoring. It can provide information on blood velocities, wall pulsatility, pulse wave velocity, and the development of atherosclerosis [1]–[5]. Three-dimensional (3D) ultrasound offers improved visualization of the 3D anatomy and volumetric flow of the CA, potentially enabling more accurate plaque volume estimation, stenosis grading, and characterization of the CA wall mechanics [5]–[10]. Furthermore, 3D image acquisition may alleviate the need for a highly skilled ultrasonographer to select the proper imaging planes within the complex 3D structure of the CA [11]. Also, such 3D capabilities would be beneficial for a wearable device for long-term monitoring.

However, to achieve wide-view high-resolution the CA 4D images, a fully populated matrix transducer with a relatively large footprint is needed. In conventional technology, such an array would require the element pitch of a corresponding linear array in both dimensions, requiring many thousands of elements in total [12], [13]. This would require channel

This work was supported by the project TOUCAN (with project number 17208) of the OTP research programme which is financed by the Dutch Research Council (NWO).

Yuyang Hu, Antonius van der Steen and Johannes Bosch are with the Department of Cardiology, Erasmus MC University Medical Center, 3000 CA Rotterdam, The Netherlands (e-mail: y.hu.@erasmusmc.nl).

Didem Dogan and Geert Leus are with the Department of Micro Electronics, Delft University of Technology, 2628 CJ Delft, The Netherlands.

Michael Brown and Pieter Kruizinga are with the Department of Neuroscience, Erasmus MC University Medical Center, 3000 CA Rotterdam, The Netherlands.

Pieter Kruizinga is also with the Department of Micro Electronics, Delft University of Technology, 2628 CJ Delft, The Netherlands, and the Department of Neurosurgery, Erasmus MC University Medical Center, 3000 CA Rotterdam, The Netherlands.

Mahé Bulot, Maxime Cheppe, Guillaume Ferin, are with the Active Probe Group, Innovation Department, Vermon SA, Tours, France.

***Highlights***

- **A cUSi system aimed at carotid artery imaging was designed and fabricated, using a custom 240-element matrix probe with a large 40 × 24 mm² aperture and a spatial encoding mask.**
- **Large volume cUSi imaging with high resolution and high frame rate was shown to be feasible with our system and computational reconstruction.**
- **This work shows that cUSi allows large-volume 4D imaging with simple and cost-effective hardware.**

reduction in the probe (e.g., by built-in electronics) and/or a very complex imaging system [13], [14]. Application-specific integrated circuits offer a solution to reduce the system complexity [15], [16]. The number of channels can be decreased by performing sub-aperture beamforming within the probe, or multiplexing element signals, both at the cost of framerate. An alternative solution, such as a sparse array [17], [18], can provide a trade-off between channel count and image quality by arranging elements sparsely across the entire probe aperture to minimize background clutter level. Using row-column addressing on matrix arrays effectively reduces the required number of channels from N×N to N+N [19], [20]. row-column addressing arrays provide a larger active aperture and improved penetration depth, at the cost of more complex transmit schemes and beamforming methods. Their element size and pitch are still tied to the ultrasound wavelength to meet the Nyquist criterion for spatial sampling.

Developments in computational ultrasound imaging (cUSi) provide a radically different way to achieve simpler transducer designs at the cost of higher computational complexity. A transducer for 3D imaging can be constructed with only a few large-pitch elements, covered by an aberration mask to create a spatially encoded acoustic field. This field provides a specific signature to signals scattered from different locations of the image. By characterizing these signals and using them in a computational inversion approach, a high-resolution reconstruction can be achieved from a spatially highly undersampled acquisition. This method was previously demonstrated in 3D volumetric imaging reconstruction with a single-channel transducer [21], 3D scanning acoustic microscopy [22], and 4D mouse brain hemodynamics [23].

In the context of application to the CA, we aim to design a cUSi system with relatively few elements that can provide us with reliable 4D B-mode images and Doppler blood flow information of the CA. We designed and constructed a custom transducer suitable for combination with an aberration mask to realize a computational imaging approach for CA. In this article, we present our system design, fabrication, and associated image reconstruction (Section II). The system was characterized and calibrated by hydrophone measurements, and its volumetric B-mode imaging performance was validated in a phantom study, including high-frame-rate transmission using two different reconstruction methods. All results are presented in Section III. Discussion of the results and limitations of this study are presented in Section IV, followed by conclusions (Section V).

## II. Methods

### A. Computational ultrasound imaging

In cUSi, we aim to reconstruct ultrasound images from fewer sensors than traditional imaging by employing computational approaches. In traditional ultrasound imaging, the backscattered sound field is sampled densely (according to Nyquist criteria in time and space), and the images (representing the spatial scatterer distribution) are reconstructed based on wave travel time and sensor geometry, typically delay-and-sum (DAS) [24], [25]. Arrays of ultrasound elements with a pitch determined by the ultrasound wavelength are employed. For imaging large superficial structures like the CA in 3D, a classical approach would thus require many thousands of ultrasound elements, and result in a highly complex and costly solution. When sampling the same field of view with a much lower number of sensors, we face severe undersampling issues.

These can be partly overcome by employing a form of spatial coding [21], as explained in Fig. 1. When imaging with a traditional pulse-echo approach (Fig. 1a), echoes of pulses from scatterers at the same depth exhibit highly similar pulse-echo responses (PERs), which precludes separation of the source signals from the channel radio-frequency (RF) data, making it more difficult to accurately localize scatterers within the field of view. To address this issue, a randomized aberration mask is placed in front of our large elements (Fig. 1b). This mask is fabricated from materials with sound speeds distinct from those of water or tissue. As acoustic waves pass through the mask, spatially varying delays, diffraction and attenuations are introduced, creating a complex, spatially encoded acoustic field. Ideally, this field will have a unique pulse signature for each spatial position, and scatterers at different locations exhibit unique PERs. By incorporating these unique responses into reconstruction, even when echoes from multiple objects are combined in a limited number of channel RF signals, they remain separable and can accommodate high-spatial-resolution reconstructions.

#### 1) Linear system model

Because of the complexity and spatial heterogeneity of this encoded acoustic field, conventional beamforming methods that only rely on the geometry of the grid and elements, such as DAS, are unsuitable. Model-based reconstruction methods are employed as alternatives.

Specifically, our cUSi approach employs a linear model $\boldsymbol{y} = \boldsymbol{A}\boldsymbol{x}$ to describe our imaging system. In this model, $\boldsymbol{y}$ denotes the vector of RF signals as received by each element, and $\boldsymbol{x}$ denotes each voxel's backscattering intensity within the volumetric region of interest (ROI). The system matrix $\boldsymbol{A}$, describing the contribution from a given $\boldsymbol{x}$ to the response $\boldsymbol{y}$, is a very large matrix that encapsulates the PER per element for every voxel in the ROI for each transmission.

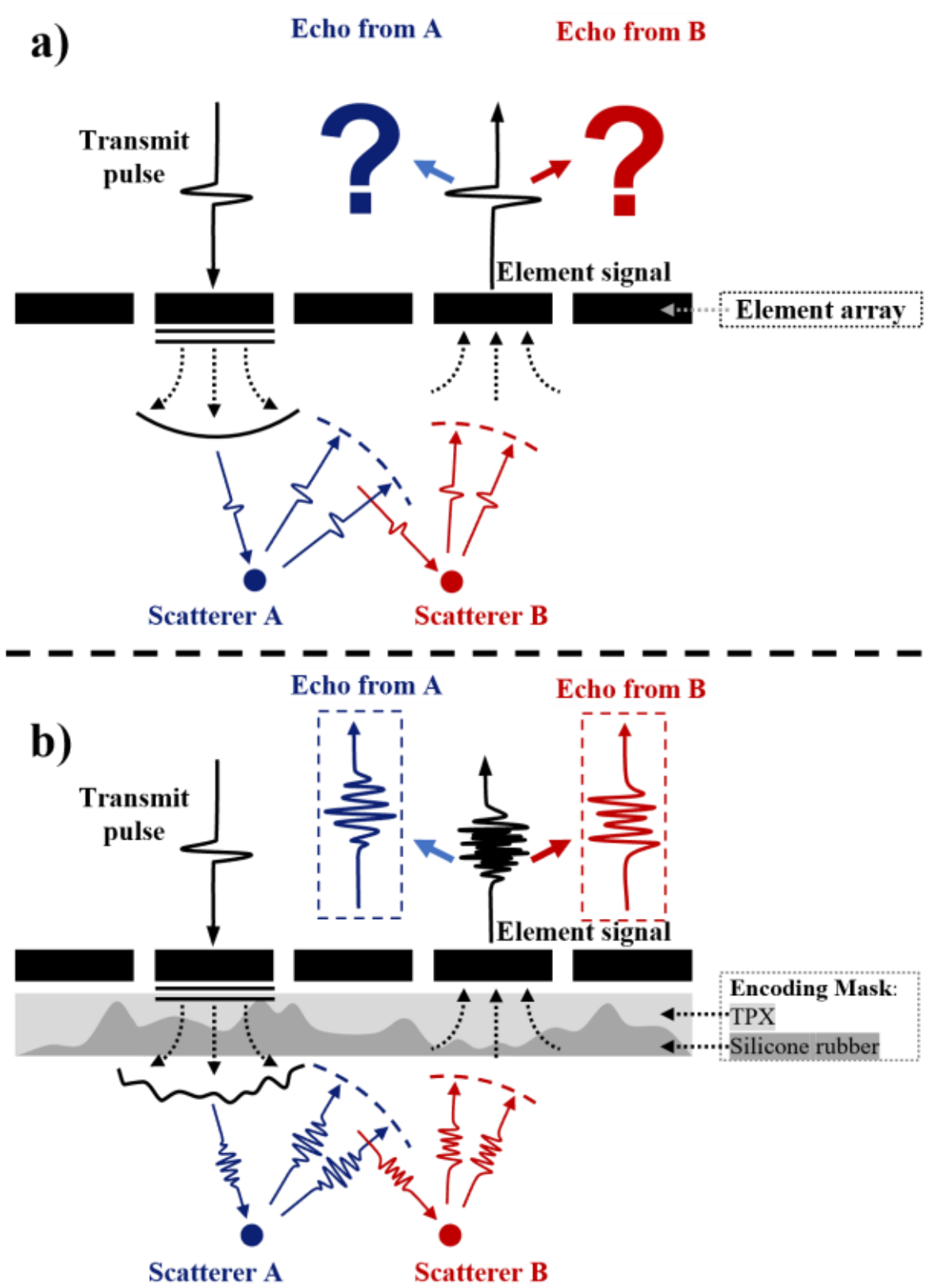

Fig. 1 Principle of spatial encoding for computational ultrasound imaging. a) Standard array: Pulse echo responses (PERs) from different positions/scatterers are difficult to distinguish in the received element signals. b) Element array with encoding mask. The mask creates unique PERs per location that can function as signatures and facilitate the localization of the scatterers in computational reconstruction.

Constructing a precise system matrix $\boldsymbol{A}$ is crucial for accurate cUSi reconstruction. In this work, we construct the $\boldsymbol{A}$ matrix from hydrophone measurements of the one-way (transmitted) pressure field. This allows us to capture a realistic acoustic field that includes the mask's local delay effects and secondary factors such as internal reflection and manufacturing variations. We apply the reciprocity theorem $\boldsymbol{A}_{(n_t,n_r)} = \boldsymbol{A}_{n_t} * \boldsymbol{A}_{n_r}$ [26], convolving the one-way fields of transmitting elements $\boldsymbol{A}_{n_t}$ with the receiving elements $\boldsymbol{A}_{n_r}$, to effectively model the pulse-echo responses $\boldsymbol{A}_{(n_t,n_r)}$ for different transmit schemes (e.g., synthetic aperture transmit, Hadamard, plane wave). Consequently, the resulting system matrix $\boldsymbol{A}$ better matches the actual device behavior compared to a purely simulation-based calibration, as was done in previous works [27], [28].

In total, the cUSi system model is structured as follows:

$$\boldsymbol{y}_{(n_t,\, n_r)}(\omega) = \sum_{x=1}^{N_x} \sum_{y=1}^{N_y} \sum_{z=1}^{N_z} \left( \boldsymbol{A}_{(n_t,\, n_r)}(\omega, x, y, z) \times \boldsymbol{x}(x, y, z) \right). \quad (1)$$

Here, $\boldsymbol{y}$ is represented as the sum of echoes from all voxels for each specific transmit and receive element, where a voxel echo is given by the product of the pulse-echo response $\boldsymbol{A}$ and the backscattering intensity $\boldsymbol{x}$. Note that $\boldsymbol{A}$ and $\boldsymbol{y}$ can be expressed either in the frequency domain ($\omega$) or the time domain ($t$) in this system model. In our study, we chose the frequency domain for a convenient frequency down-sampling during reconstruction and easier calculations of the PERs.

## B. System design and fabrication

### 1) 240-element matrix probe

Our CA imaging application aims to achieve 4D volumetric imaging with a low number of elements, maximally 256, to maintain compatibility with our Vantage 256 research ultrasound system (Verasonics Inc., Kirkland, WA, USA). A footprint suitable for the desired field of view and a suitable center frequency for the CA imaging was chosen. To find a compromise between field of view and spatial sampling density, we refer to our prior configuration study of a cUSi emulation system [28], which achieved high-resolution 2D CA imaging with a 12-element linear array of 40 mm aperture and 10-wavelength pitch (at 5 MHz). Extending this linear array configuration to 2D, we settled for a 240-element regular matrix probe (20 × 12 arrangement) with an element pitch of 2 × 2 mm (Fig. 2a). This configuration achieves a footprint of 40 × 24 mm², well-suited for the CA imaging, at a pitch of ~8 wavelengths (~6 MHz). It provides 12-20 spatial sampling points per dimension, which is expected to provide a reasonable reconstruction similar to the 2D case. A regular square element matrix design was chosen for easier manufacturability, and to maximize element size for maximal sensitivity. This configuration was also validated through a 5 × 5 array simulation in k-Wave before fabrication [27].

We designed our matrix probe in collaboration with our industrial partner Vermon (Tours, France), who also fabricated it. As depicted in Fig. 2a (bottom), a 48 × 32 mm$^2$ slice of 1-3 connectivity piezo-composite metallized on both sides was diced into a 24 × 16 elements array, leaving two rows of elements along each side as a margin. The active 240 elements were wired individually via a flexible printed circuit board on the back and grounded on the front. Detachable, exchangeable aberration masks should be acoustically coupled directly to the probe surface. To allow this, a flat soda-lime glass layer was used as the outer matching layer on the piezo-composite (Fig. 2a bottom). The thickness of the glass matching layer was chosen as $\lambda/2$ (~ 500 µm) to maximize energy output (based on KLM simulations [29]) while ensuring sufficient mechanical stability.

At the bottom of the acoustic stack is a tungsten-epoxy damping layer supported by an aluminum framework that also serves as a heat sink. The whole is encapsulated in an ABS shell and interconnected to a GE-type ZIF connector for coupling to the Verasonics Vantage-256 system.

### 2) Encoding mask

The encoding mask is designed to achieve sufficient acoustic delays with minimal energy loss while remaining thin and flat on both sides to ensure easy acoustic coupling with the probe surface and the neck without air bubble entrapments. In this work, we introduce a dual-layer mask (Fig. 2b). The encoding pattern (Fig. 2c) defines the varying local thicknesses of the bottom layer which is covered by the top layer up to the flat outer surface. Specifically, we selected TPX and silicone rubber as our layer combination because they enable a smooth transition of acoustic impedance from the piezo-composite to human tissue (approximately 20 → 13 → 2 → 1→ 1.5 MRayl), along the acoustic path (red arrow, Fig. 2d bottom). The desired mask pattern was CNC-milled into a slice of TPX, leaving a protruding border around the pattern edge that was filled with a degassed silicone rubber solution (KE-109E-A/B, Shin-Etsu Silicones Europe B.V., Almere, the Netherlands), specifically chosen for its low acoustic attenuation (0.12 dB/mm/MHz) and reliable adhesion to TPX. Its sound speed (1101 m/s) is lower than water (1480 m/s), while that of TPX is higher (2190 m/s), which enables the same delay variation in a thinner mask compared to using only TPX.

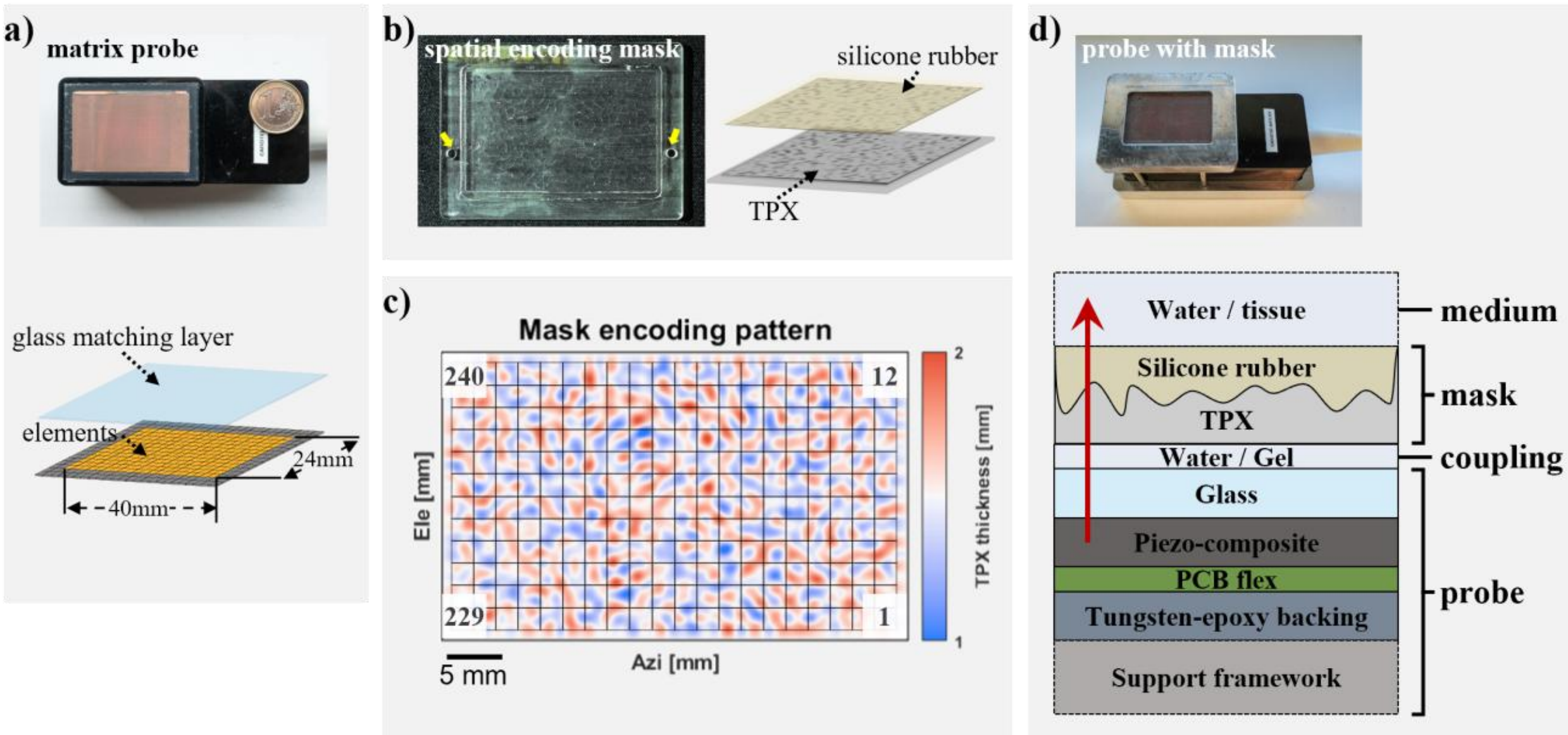

Fig. 2 Realized cUSi system and its components. a) 240 elements matrix probe (top photo) with an active aperture of 40×24 mm$^2$ consisting of 20×12 elements and a glass matching layer; b) Spatial encoding mask (left, transparent), consisting of a patterned slice of TPX and a layer of silicone rubber. Yellow arrowheads indicate alignment holes. c) Varying thickness pattern of TPX layer for encoding; element indices at the four corners are annotated for reference. d) Mask coupled to the probe by an aluminum frame (top); acoustic stack (bottom), layer thickness not to scale; the red arrow indicates the acoustic path.

The mask encoding pattern shown in Fig. 2c was defined on an 800 × 480 grid using uniform random values at a resolution of 50 µm. Spatial smoothing was applied to limit internal reflections (local gradient < 54° based on a Snell's law calculation). The resulting 2D thickness distribution spans roughly 0–1 mm, or ~3 wave periods of delay. An extra 1 mm layer of TPX or silicone rubber was added on each side of the mask to ensure structural stability, leading to a total thickness of 3 mm.

An aluminum fixation frame (Fig. 2d top) was used to clamp the edge of the mask to the probe and couple them through a thin layer of ultrasound gel (Parker Laboratories, Inc., Fairfield, NJ, USA). To facilitate repeatable attachment and detachment, the inside of the top fixation part has two pins that align with the two holes on the edge of the mask (Fig. 2b, yellow arrowheads). The probe fits tightly within the framework, and the framework with the mask can be attached and detached using four screws. This ensures that the mask and probe are tightly and reliably aligned each time.

### 3) *Probe and field characterization*

The probe was characterized by hydrophone measurements, and the acoustic field of individual elements (without and with the mask) was measured to build the system model (calibration). A 0.2 mm needle hydrophone (Precision Acoustics Ltd. Dorchester, UK), mounted on a 3D stepper motor stage, was positioned 8 mm from the mask surface (far field for a single element) in a water tank. The hydrophone then scanned a 55 × 40 mm² area at this depth, coaligned with the probe aperture, with a step size of 0.1 mm. The respective elements of the probe were driven by a 20 V, 30 ns excitation pulse, and the hydrophone signals were sampled with a 250 MHz sampling rate by a programmable digitizer (M4i.4450-x8, Spectrum, Germany). A Hadamard-encoded synthetic transmit aperture scheme [30] was implemented to boost the signal-to-noise ratio (SNR) with respect to single-element excitation. Since the number of elements is not an exact power of two, we trimmed the original 256 × 256 Hadamard matrix to 240 × 256 by assigning only the first 240 rows to a corresponding element. All 256 transmissions were then executed accordingly. Despite this slight dimension mismatch, we could still decode each element's response with negligible artifacts from the incomplete Hadamard matrix. After decoding, the pressure field for each element at the scan depth was obtained. We defined the element impulse response as the temporal response measured at the pixel exhibiting the maximum amplitude in the no mask case.

The same hydrophone scanning was performed for the probe coupled to the mask. Using the measured pressure field at this specific depth, without and with the encoding mask, we calculated the volumetric acoustic field by applying the angular spectrum method [31], [32]. This approach yielded a manageable total scanning time of about 10 hours for the single scanning plane.

## C. Reconstruction methods

### 1) Matched filtering

The system matrix established by the calibration process enables us to solve the reconstruction problem using model-based methods. However, since the matrix $\boldsymbol{A}$ encompasses a very large number of voxels (on the order of $10^6$), vastly exceeding the number of available measurement samples ($N_x N_y N_z \gg N_\omega N_t N_r$), the resulting system matrix becomes extremely large and ill-posed. As a result, a direct matrix inversion $\boldsymbol{A}^{-1}$ is infeasible. In this study, one of the approaches we adopted to address this challenge is the matched filtering (MF) method [33], [34]:

$$\hat{\boldsymbol{x}}_{MF} = \sum_{n_t=1}^{N_t} \sum_{n_r=1}^{N_r} \left( \boldsymbol{A}^{\boldsymbol{H}}_{(n_t,\, n_r)} \times \boldsymbol{y}_{(n_t,\, n_r)} \right). \tag{2}$$

Instead of computing the inverse of $\boldsymbol{A}$, MF utilizes the Hermitian (conjugate transpose) of the system matrix as the pseudoinverse. This method is computationally efficient, requiring only modest computational resources and execution time. It aligns well with the underlying physics of wave propagation, providing a robust estimation $\hat{\boldsymbol{x}}_{MF}$. However, MF is sensitive to the non-uniformity of the encoded acoustic pressure field, which can lead to inhomogeneous intensity distribution in the reconstructed image, as reported in [28].

### 2) Least-squares with QR decomposition

To further improve reconstruction quality, we additionally employed the least-squares with QR decomposition (LSQR) method [35]:

$$\hat{\boldsymbol{x}}_{LSQR} = \operatorname{argmin}_{\boldsymbol{x}} \left\| \sum_{n_t=1}^{N_t} \sum_{n_r=1}^{N_r} \left( \boldsymbol{A}_{(n_t,n_r)} \hat{\boldsymbol{x}} - \boldsymbol{y}_{(n_t,\, n_r)} \right) \right\|_2^2. \tag{3}$$

This algorithm minimizes the global residual between the estimated projection $\boldsymbol{A}_{(n_t,n_r)}\hat{\boldsymbol{x}}$ and the measurement $\boldsymbol{y}_{(n_t,\, n_r)}$ across all transmit-receive pairs. The $\hat{\boldsymbol{x}}_{MF}$ is used as the initial estimation, and the estimated $\hat{\boldsymbol{x}}_{LSQR}$ is iteratively refined to better approximate the true $\boldsymbol{x}$.

Compared to MF, LSQR makes more comprehensive use of the system matrix and can effectively solve the inhomogeneous intensity distribution in MF reconstructions. However, this method is computationally more demanding and susceptible to noise fitting, particularly if there is a mismatch between the actual system and the calibrated system matrix. To balance reconstruction quality and computational efficiency, we empirically employed 5 iterations for our final reconstruction result and evaluated its performance.

### 3) Delay-and-sum

Additionally, for the non-encoded case (without the mask), we performed direct reconstruction using the built-in DAS beamformer provided by the Verasonics system, serving as a control reconstruction.

## D. Data acquisition and preparation

We validated the 3D B-mode performance of our system on a general-purpose tissue-mimicking phantom (CIRS model 040GSE, Norfolk, VA). The probe was driven by a 3 MHz, 2-cycle tone burst with a pulse repetition frequency of 500 Hz. The low transmit frequency was chosen for observed increased reconstruction performance, likely because of the relatively better spatial sampling. Acquisitions were performed over a depth range of 5-55 mm, achieving an ROI size of 40×24×50 mm$^3$.

Two transmit schemes, Hadamard-encoding and plane-wave compounding, were applied. We expect the Hadamard-encoding scheme to provide the best reconstruction quality since it provides more measurements with good SNR, which scheme is as we described in Section II-B3 for the field calibration (240 elements×256 transmit). Plane-wave compounding is considered more practical for the CA imaging since it can achieve higher volume rates. Sixteen angled plane waves were applied, uniformly distributed in 4 steps per direction and steered within ±12° azimuth and ±7° elevation. These steering angles ensured that at 50 mm depth, 84% of the ROI was covered by at least 8 plane-waves.

RF data were acquired using the Verasonics system at a sampling rate of 25 MHz. To reduce the high memory demands of model-based reconstruction, both the RF data and the system matrices were sub-sampled in the frequency domain. A frequency band of 2.4-5.0 MHz (200 samples) was selected for processing.

We performed Hadamard decoding on the RF data before applying the built-in DAS reconstruction in the Verasonics system. In contrast, for our model-based MF and LSQR methods, this decoding process was automatically handled within the system matrix.

## E. Evaluation

In this study, we evaluated our cUSi system reconstruction for different configurations. For clarity, we will refer to the case of imaging without a mask as “no encoding” (or “non-encoded”) and the case with the mask as “encoding” (or “encoded”). Using the Hadamard transmit scheme, we compared the B-mode reconstructions of non-encoded DAS, non-encoded MF, and encoded MF. The LSQR was not applied here because of the extreme computing time. For the 16-angle plane wave scheme, we compared non-encoded DAS and both MF and LSQR in the non-encoded and encoded cases. The matrix probe was repositioned before and after mounting the mask, resulting in a slight offset of the ROI between the acquisitions. To ensure a fair comparison, we manually applied minor rotation and translation adjustments.

The resolution performance of the system was assessed using the point spread function (PSF) of the wire targets in the ROI. As illustrated in Fig. 5b, all wire targets in sub-ROIs 1 and 2 were used to evaluate lateral and axial resolution, with the full width at half maximum (FWHM) measured for each. For better visualization, maximum intensity projections (MIPs) along the azimuth dimension were shown for wire targets in sub-ROI 1 (at the same depth), and MIPs along the depth dimension were shown for wire targets in sub-ROI 2 (at varying depths). Contrast performance was represented by calculating the contrast ratio (CR) between two inclusions (inc 1 & 2 in Fig. 5b) and a background region (BK in Fig. 5b) at the same depth. The CR was defined as:

$$\mathrm{CR} = 20 \times \log_{10} \left( \frac{\mathrm{rms(inc)}}{\mathrm{rms(BK)}} \right). \tag{4}$$

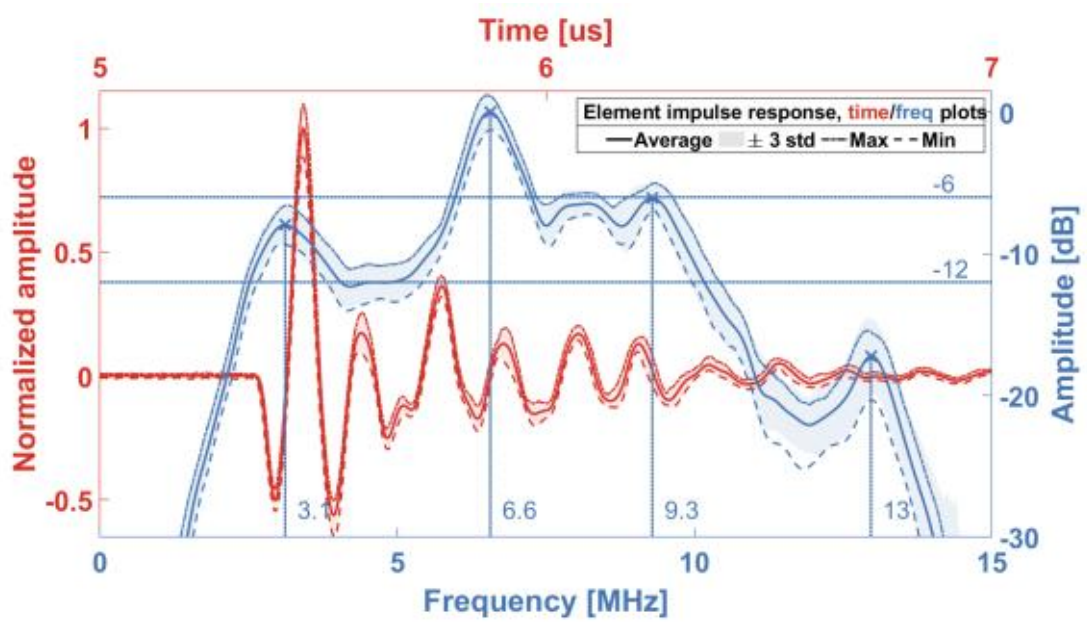

Fig. 3 Impulse responses over all 240 elements of the matrix probe, normalized to the peak of the mean element response. Average (solid line) with ±3 standard deviations (shaded band), maximum (dash-dot line) and minimum (dashed line) across all elements. Red plots represent the temporal domain; blue plots represent the frequency domain.

Here rms(·) corresponds to the square root of the mean of the squared intensities within the respective area.

We observed that some wire targets could not be reliably reconstructed in certain cases and might introduce bias into the evaluation. To ensure consistency, the topmost and deepest wire targets were excluded from the quantitative FWHM analysis.

## III. Results

### A. Probe and field characterization

Fig. 3 shows the range of impulse responses of all 240 elements, normalized to the peak of the mean element response. The average temporal and frequency responses of all elements are shown, with a band indicating three standard deviations, and the maximum and minimum responses across all elements. No element exhibited abnormal temporal or spectral behavior. The amplitude of the weakest element response (in the temporal domain) was only 2.6 dB lower than the strongest. This implies that all 240 elements functioned properly and exhibited highly consistent performances. The matrix probe has a peak frequency of 6.6 MHz and a relatively narrow -6 dB bandwidth of 12%. However, two additional response peaks are observed at 3.1, and 9.3 MHz, likely resulting from the use of the 500 µm glass matching layer. When evaluated at the -12 dB level, the effective bandwidth covers 166%, allowing broader spectral utilization. Once the mask is applied, the uniformity of the element responses is intentionally disrupted; although the impulse response shape is not altered that much, the interactions across encoded elements introduce the spatiotemporal decorrelation required for cUSi reconstruction. Supplementary Fig. S1 provides a comparison of three randomly selected element impulse responses (Fig. S1a-b) and forward plane-wave fields at three positions (Fig. S1c-d) of different depths with and without the mask to illustrate these effects. Variations between element signals (both shape and peak amplitudes) only slightly increased after adding the mask. Yet, the mask caused an obvious difference in the spatiotemporal responses in the forward plane wave fields).

Fig. 4 displays the pressure field profiles for different configurations. Fig. 4a shows the pressure field generated by a single element from the surface of the mask to 55 mm depth; as expected, the large element pitch results in a highly directional acoustic field. When applying the encoding mask (Fig. 4d), the field diverges significantly—for example, the single-element opening angle at 25 mm depth increases from 11.5° to 44.5° (−12 dB cutoff). This implies that during reconstruction, signals from more adjacent elements can contribute to each voxel, enhancing reconstruction quality.

Fig. 4b and 4e present a plane wave field of the probe without and with the mask. The mask significantly disrupts the homogeneity of the acoustic field, effectively enabling spatiotemporal encoding. However, we also observe a decrease in penetration depth, which might be caused by the reduced wavefront coherence and additional attenuation introduced by the mask materials.

The acoustic fields at 25 mm depth (Fig. 4c & f, top left) and their corresponding system matrices were further analyzed. The k-space distribution across the entire imaging plane at this depth (Fig 4c & f, top right) reveals a more comprehensive spatial frequency coverage with higher-frequency components in the encoded field, which are needed to obtain a higher spatial resolution. This assumption is further supported by the narrower correlation peak between the central pixel's PER and its surrounding pixels in the encoded system matrix (Fig. 4f, bottom) compared to the non-encoded case (Fig. 4c). Such correlation patterns are commonly used to approximate the system's theoretical local PSF [23]. Additionally, the randomized encoding pattern of the mask effectively suppresses the secondary correlation peaks that appear along the azimuth and elevation directions in the non-encoded system (Fig. 4c). These secondary peaks, arising from the periodic element layout and insufficient spatial sampling, are typically associated with grating lobes in imaging and can lead to ambiguous scatterer localization.

In our mask design, we aimed to minimize transmitted energy loss by selecting materials with smooth acoustic impedance transitions and relatively low attenuation. As a result, the encoded field shows only a ~3 dB reduction in time-integrated field power and a ~6.6 dB reduction in peak intensity compared to the non-encoded case. Nevertheless, it should be noted that this loss is an inherent trade-off of the encoding scheme and will contribute to reduced image contrast.

### B. Phantom validation

#### 1) Hadamard transmit

Fig 5 demonstrates the in vitro volumetric B-mode imaging performance of our cUSi matrix system using the Hadamard transmission scheme. Among the three reconstruction approaches, DAS, MF without encoding, and MF with encoding, the encoded MF reconstruction (Fig. 5a, right) yields the best image quality. All wire targets are clearly and sharply resolved, and Inclusion 1 (Fig. 5a, red regions) appears as a well-formed, natural-looking circle close to the ground truth. While the non-encoded MF (Fig. 5a, middle) shows worse resolution and less accurate inclusion shape, it still successfully reconstructs all targets. Notably, this configuration achieves the best CR with lower background energy. As a result, Inclusion 2 (Fig. 5a, blue regions), which has inherently lower contrast, remains identifiable in the MIP image. The DAS reconstruction (Fig. 5a, left) performs the worst. The topmost wire target is not discernible and pronounced grating lobes are observed around the reconstructed wire targets in ROI 1. Besides, Inclusion 1 is only roughly localized, with no clear lateral boundary. A comparison of the different 3D reconstructions is shown in Supplementary Movie M1. An image corresponding to Fig. 5a without the boundaries of inclusions 1 and 2 overlaid is provided as Supplementary Fig. S2a to facilitate the reader's interpretation.

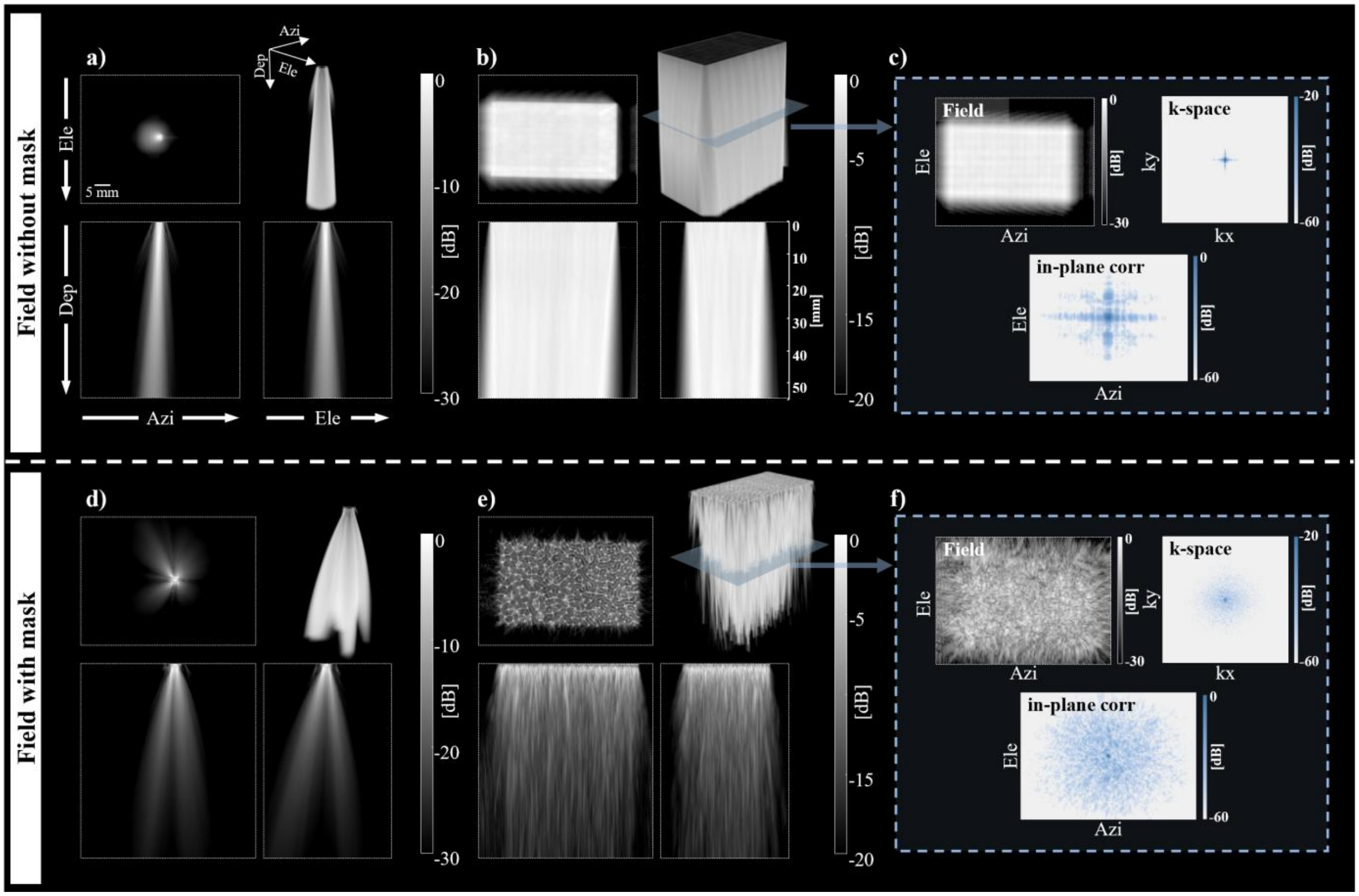

Fig. 4 Acoustic field of single-element and plane wave excitation with and without mask, from the surface of the mask to 55 mm depth, each shown as a 3D view (top-right) and MIP along three dimensions. a) Single-element field without mask. b) Plane-wave field without mask. c) cross-sectional non-encoded plane-wave field profile at 25 mm depth (blue plane in b)), its k-space and in-plane correlation of PER. d) Single-element field with mask; e) Plane-wave field with mask; f) cross-sectional spatial encoded plane-wave field profile at 25mm depth (blue plane in e)). After applying the encoding mask, the single-element acoustic field d) exhibits an increased opening angle, and the plane-wave field e) reveals a more complex spatial pattern. The plane-wave encoded field profile at 25 mm in f) demonstrates broader k-space coverage compared to the non-encoded field in c), indicating more informative spatial frequency sampling. Additionally, the narrower in-plane correlation peak (c&f, bottom) implies a potentially improved spatial resolution achieved by encoding.

Quantitative analysis (Fig. 5c, Table I) shows that lateral resolution improves approximately twofold with encoded MF ($0.91 \pm 0.13$ mm) compared to non-encoded MF ($1.67 \pm 0.25$ mm) and DAS ($1.91 \pm 0.56$ mm). CR significantly improves when using non-encoded MF over DAS, especially in inclusion 1 ($8.16 \rightarrow 12.93$ dB). However, the encoding slightly reduces the CR ($12.93 \rightarrow 12.57$ dB). Axial resolution is moderately improved by MF ($1.79 \rightarrow 1.16$ & $1.22$ mm), particularly for deeper wire targets (Fig. 5d). Full quantitative results are provided in Table I.

A comparison between the lower-frequency band (2.4-5.0 MHz) and a higher-frequency band (5.3-7.9 MHz) MF reconstructions is presented in Supplementary Fig. S3. Although the probe has been excited at 3 MHz and 6.6 MHz, respectively, to optimize the SNR for each reconstruction, the image quality at the higher frequency remains heavily degraded, especially with encoding.

### 2) 16-angle plane wave transmit

Fig. 6 shows reconstructions using the 16-angle plane wave transmission scheme, where only azimuth-depth MIPs are displayed for visualization. Similar to the Hadamard case, encoded MF outperforms non-encoded MF and DAS in terms of resolution and reconstruction of Inclusion 1 (Fig. 6a, bottom middle). However, the deepest wire target appears only vaguely, and the reconstructed image showed a non-uniform intensity distribution. In the non-encoded MF case (Fig. 6a, top middle), CR remains high, and all wire targets are clearly visible, apart from stripe-like artifacts that emerge in the near-field. In the DAS case (Fig. 6a, top left), severe undersampling artifacts are observed, with strong striping and grating lobes, particularly around the near-field wire targets.

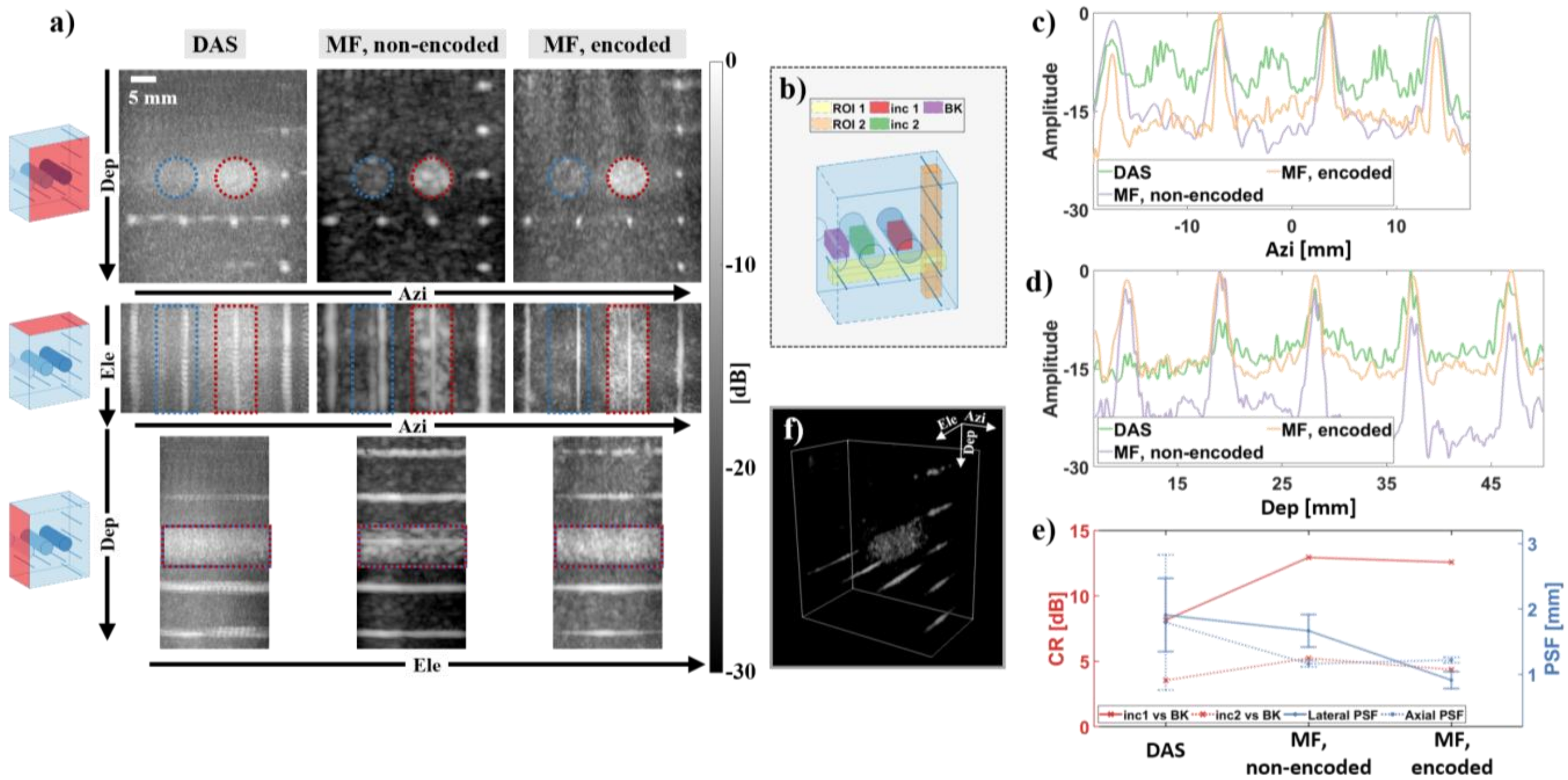

Fig. 5 Comparison of volumetric B-mode reconstructions using a Hadamard transmit scheme. Reconstructions are visualized as MIP along three dimensions. a) From left to right: B-mode images reconstructed by DAS, non-encoded MF, and encoded MF. Top to bottom: MIP projections in elevation, depth, and azimuth directions. The boundaries of inclusions 1 (red regions) and 2 (blue regions) are overlaid for clarity. The same image without these boundaries is shown in Supplementary Fig. S2a. c) Lateral PSF, evaluated from azimuth-MIP of wire-targets within ROI 1 (yellow box in b)); d) axial PSF evaluated from depth-MIP of wire-targets within ROI 2 (orange box); e) CR between inclusion 1 (red box) and background (BK, purple box; red solid line), and between inclusion 2 (green box) and background (red dashed line), and averaged PSFs: lateral (blue solid line) and axial (blue dashed line).The encoded MF yields the best lateral resolution, whereas the non-encoded MF provides the highest CR. DAS exhibits the worst PSF and CR. f) Encoded MF reconstruction shown in 3D view. The numerical results of c-e) are reported in Table I.

Applying LSQR generally enhances image quality considerably, particularly by improving resolution and suppressing artifacts. In the non-encoded LSQR reconstruction (Fig. 6a, top right), PSFs are visibly sharper than in the MF case, and near-field striping is largely suppressed. Both Inclusion 1 (Fig. 6a, red regions) and Inclusion 2 (Fig. 6a, blue regions) are reconstructed with greater structural completeness. In the encoded LSQR case (Fig. 6a, bottom right), improvements mainly appear as more uniform intensity distribution and modest PSF enhancement. However, LSQR also increases the background/clutter level in this case; the regions beside Inclusion 1 appear significantly brighter, and the vaguely seen deepest wire target is completely obscured. A comparison of all five 3D reconstructions is shown in Supplementary Movie M2. An image corresponding to Fig. 6a without the boundaries of inclusions 1 and 2 overlaid is provided in Supplementary Fig. S2b to facilitate the reader's interpretation.

Quantitative results are listed in Table I. Lateral PSF (Fig. 6b) improves significantly with encoding and was further enhanced by LSQR. Compared to non-encoded MF (1.99±0.3 mm), both encoding (1.08±0.22 mm) and LSQR (1.16±0.17 mm) achieve ~2× improvement, with their combination providing the best lateral resolution (0.85 ± 0.16 mm). Although DAS also shows a relatively low lateral PSF (1.51 ± 0.62 mm), this seems to be a side effect of the strong vertical dark-stripe artifacts (Fig. 6a top left) that cut away parts of the sidelobes. As shown in Fig. 6d, the CR of Inclusion 1 increases from 9.46 dB (DAS) to 11.99 dB with non-encoded MF. However, encoding leads to a more pronounced CR drop (11.99 → 9.78 dB) compared to the Hadamard case (12.93 → 12.57 dB), and LSQR slightly reduces this further, likely due to an increased background intensity. In terms of axial resolution, MF provides minor improvements compared to DAS (1.87±0.44 mm) both without (1.43±0.1 mm) and with (1.6±0.2 mm) encoding, while LSQR

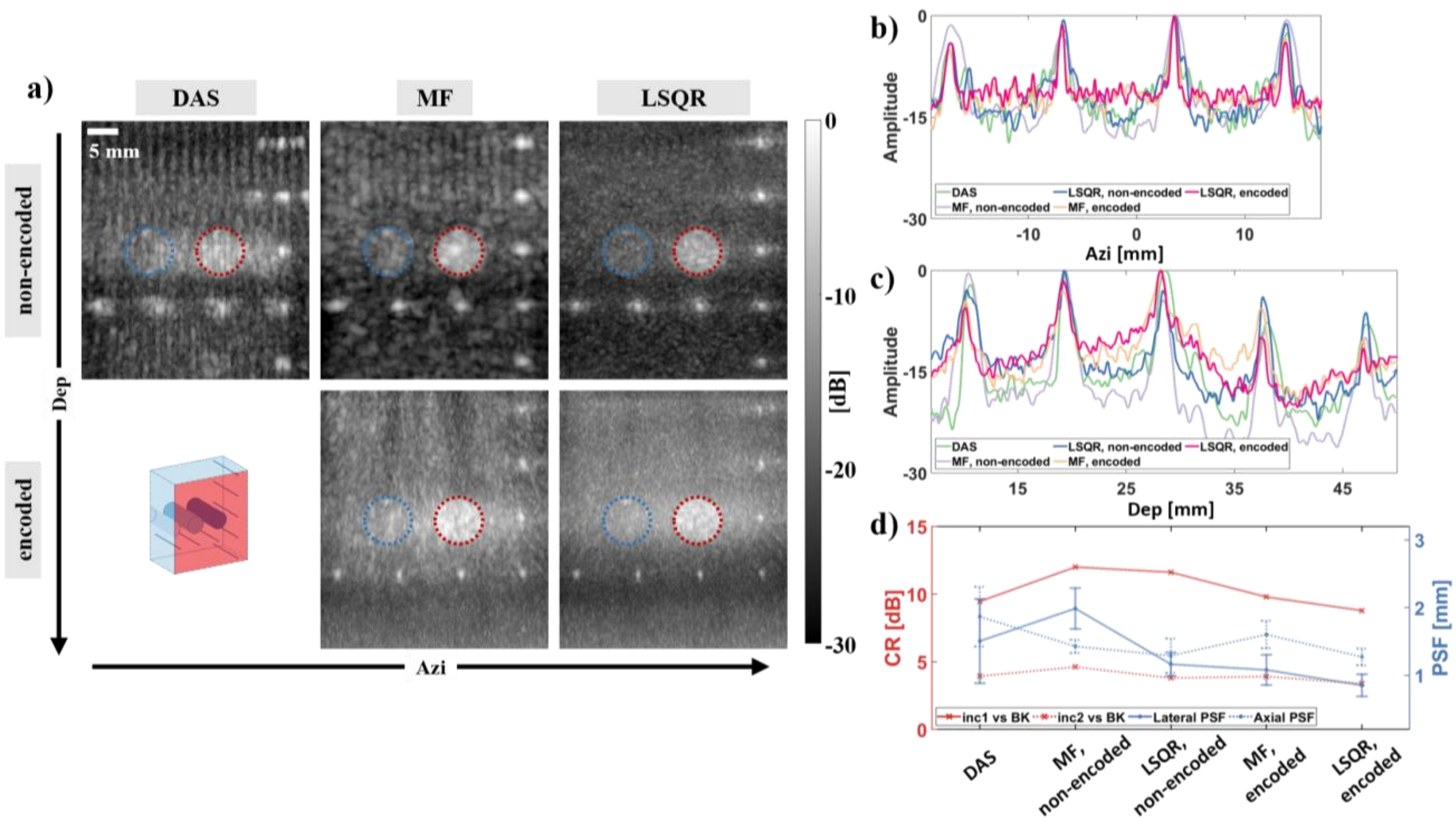

Fig. 6. a) Volumetric B-mode reconstructions using a 16-angle plane-wave imaging scheme (MIP in elevation direction). From left to right: DAS, MF, and LSQR reconstructions. From top to bottom: results without encoding and with encoding. The same ROIs, as shown in Fig. 5b, are used for PSF and CR evaluation. The boundaries of inclusions 1 (red regions) and 2 (blue regions) are overlaid for clarity. The same image without these boundaries is shown in Supplementary Fig. S2b. b) Azimuth-MIP showing lateral PSFs. c) Depth MIP showing axial PSFs. d) CR and averaged lateral and axial PSF. The lateral PSF is improved by encoding and LSQR, where the combination of both produces the best outcome. However, encoding reduces the CR, both using MF and LSQR, obscuring the deepest wire target. The axial PSF only shows minor improvement after encoding and LSQR. DAS yields the worst performance, exhibiting poor PSF and strong stripe artifacts. The numerical results of b-d) are reported in Table I.

further enhances them (1.29±0.25 mm & 1.27±0.12 mm).

## IV. Discussion

In this study, we proposed a large-footprint, low-element-number 3D cUSi system designed specifically for the CA imaging. The proposed system consists of a 240-element matrix probe with a 2 mm pitch and a specially designed aberration mask to encode the acoustic field. We characterized the probe's impulse response and the acoustic field through hydrophone measurements, which we also used to establish the system matrix. By computational reconstructions, we achieved high spatial resolution volumetric B-mode imaging over a 40 × 24 × 50 mm³ volume in a tissue-mimicking phantom. The imaging performance of our cUSi system was investigated by comparing different reconstruction methods (MF, LSQR, and DAS) under two transmission schemes (Hadamard and 16-angle plane wave), both with and without acoustic encoding.

Our results indicate that for this large-element matrix, model-based reconstruction methods (MF and LSQR) generally outperform DAS, with and without encoding (mask). DAS relies on geometric relationships between array elements and reconstruction grids, assuming each element's center as its position. However, this assumption fails to handle the local phase differences across the large element, leading to incorrect delay estimation. Model-based methods use system calibration, providing accurate RF-to-grid phase relationships. Moreover, it can tackle the directional weighting of the element field and additional encoding (Hadamard) or delays (plane waves). Thus, they yield more accurate reconstruction results compared to DAS.

In the 16-angle plane-wave case, LSQR further improved the reconstruction resolution and effectively suppressed artifacts from MF by enabling a better (in the least squares sense) fit between the data **y** and the image **x** multiplied by the model **A**. Additionally, the MF output scales directly with the varying amplitudes of the local acoustic fields, resulting in intensity variations such as near-field stripe artifacts (Fig. 6a, mid-top) and uneven background intensities (Fig. 6a, mid-bottom). However, compared to MF, LSQR is computationally intensive ( $T_{LSQR}=T_{MF}\times\#iterations\times 2$ ), and limited by its stronger dependence on the accuracy of the system matrix and on sufficient SNR. These factors may reduce or even negate the improvement over MF reconstruction in in-vivo situations, where tissue aberration and attenuation introduce some additional mismatches between the system matrix and the actual acoustic field. Further investigation will be valuable to improve imaging performance under such model distortions, e.g. through adaptive focusing methods as proposed in [36].

We also observed that acoustic field encoding (using the mask) notably improved lateral resolution in MF. This improvement is attributed to an increased variation in the PERs across transmit–receive pairs, and wider element opening angle while imaging, which results in a narrower response in the correlation of the received RF with PERs, that forms the basis of MF (2). However, this also comes at the cost of a reduced acoustic penetration and an increased background clutter, as seen in the element field profiles maps (Fig. 4a vs. Fig. 4b) and the PER correlation maps (Fig. 4c bottom vs. Fig. 4f bottom). The brighter appearance of the encoded images in Fig. 5 and Fig. 6 also reflects this elevated background clutter, indicating reduced contrast and lower effective SNR.

TABLE I
FWHM OF PSF (MEAN (±SD)) AND CR (INC1 VS BK/ INC2 VS BK) FROM PHANTOM EXPERIMENT

| | Transmit schemes | Hadamard | | 16-Plane Wave | | |
|---|---|---|---|---|---|---|
| | Reconstruction schemes | DAS | MF | DAS | MF | LSQR |
| non-encoded | Lateral PSF [mm] | 1.91(±0.56) | 1.67(±0.25) | 1.51(±0.62) | 1.99(±0.3) | 1.16(±0.17) |
| | Axial PSF [mm] | 1.79(±1.03) | 1.16(±0.05) | 1.87(±0.44) | 1.43(±0.1) | 1.29(±0.25) |
| | CR [dB] | 8.16/3.55 | 12.93/5.24 | 9.46/3.93 | 11.99/4.62 | 11.6/3.8 |
| encoded (mask) | Lateral PSF [mm] | | 0.91(±0.13) | | 1.08(±0.22) | 0.85(±0.16) |
| | Axial PSF [mm] | | 1.22(±0.04) | | 1.6(±0.2) | 1.27(±0.12) |
| | CR [dB] | | 12.57/4.37 | | 9.78/3.91 | 8.78/3.39 |

Besides, sufficient depth is needed for the element encoded fields to develop, overlap, and interfere, which affects the spatiotemporal correlation among neighboring pixels and consequently the imaging performance in terms of PSF and CR. At very shallow depths, where the element fields are less decorrelated and less overlapped, the PSF remains broad and spatially dissimilar. In deep regions, contrast becomes lower due to reduced SNR and increased clutter level from stronger field decorrelation. A reasonable trade-off between PSF sharpness and contrast is expected at the intermediate depths we target. These near-field and far-field effects are partly reflected in our results: in Fig. 5a, the reconstruction of the top wire target in the encoded MF case is incomplete in the elevation dimension (3rd column, bottom), and in Fig. 6a, the deepest wire target is barely visible compared to the others (2nd column, bottom), which indicates a loss of contrast/penetration. A direct comparison of the transmitted field distributions and corresponding CIRS phantom reconstructions of Hadamard transmission at different depths is provided in Supplementary Fig. S4 for further demonstration.

Comparing the Hadamard and 16-angle plane-wave schemes, we observed that the latter increased the frame rate 16-fold at the cost of a slight resolution reduction (~20% on average). However, encoding led to a more pronounced CR drop in the 16-angle case (Fig. 6d vs. Fig. 5e). This indicates that applying encoding might be less effective for low-contrast targets or low SNR conditions, suggesting potential benefits from combining our encoded MF with contrast-enhanced ultrasound techniques.

These observations on the effect of encoding on reconstruction quality align well with our previous emulation studies [28], where encoding significantly improved lateral resolution, LSQR further enhanced image resolution (particularly in the non-encoded case) and helped to correct intensity distribution artifacts seen with MF. In that work, encoded cUSi emulation achieved 2D CA imaging with a lateral resolution of $0.53 \pm 0.13$ mm, axial resolution of $0.99 \pm 0.17$ mm, and a CR of 9.22 dB. Although our 3D real device shows slightly lower resolution ($0.87 \pm 0.12$ mm lateral, $1.27 \pm 0.09$ mm axial), it provides improved CR (12.19 dB), implying the applicability of cUSi in realistic 3D imaging scenarios.

We would like to stress that better reconstruction quality was observed when using a frequency band lower than the probe's peak frequency, as shown in the supplementary Fig. S3. Although minor sacrifices in lateral resolution were noted, this strategy effectively increased the contrast ratio. This improvement is likely due to the alleviation of spatial undersampling, as the element pitch becomes effectively ~4λ at lower frequencies. Improved acoustic penetration and reduced mask attenuation at these frequencies further contribute by providing better SNR. In addition, the mask introduces less phase delay for lower frequencies, limiting the clutter level at the cost of a slight spatial resolution loss. We observed this effect of delay increase in our previous study [28]. These observations highlight the necessity of considering frequency-band selection in cUSi and warrant further investigation in future work.

The probe characterization showed an unconventional frequency response, which may be associated with the adopted half-wavelength glass matching layer. This design is intended to provide high energy output at the desired working frequency (6 MHz) while also offering more mechanical strength to the probe's outer surface for mounting the mask. While such a frequency response is not a principal limitation for model-based reconstruction, it may still introduce additional inconvenience in transmission, reception, and post-processing by restricting the usable frequency range. A quarter-wavelength matching layer design may therefore be reconsidered to achieve a smoother and broader frequency response.

In general, our cUSi system provides a solution for achieving high-resolution imaging with a large aperture while maintaining a simple hardware design, something that is not easily achieved with approaches like sub-aperture beamforming or sparse array probes [16], [18]. As an alternative, row-column addressing has also been proposed for large-aperture imaging with a reduced number of read-out channels [20]. However, the ROI in cUSi is not strictly limited by the aperture size or the natural focusing of the elements. Moreover, cUSi can accommodate more flexible transmit and receive schemes, such as divergent wavefields for a larger field of view or converging wavefields, which allow a more 'focused' transmission to a certain depth; however, a systematic evaluation of such schemes is beyond the scope of this paper.

Finally, this study is still limited to validating our system in *in vitro* phantom experiments. Based on these results, we anticipate sufficient performance for the CA B-mode and flow measurements, which will take place in the next phase of research. The 16-angle plane-wave approach demonstrates reasonable image quality at higher frame rates (effectively 500 Hz). Non-encoded reconstruction remains a promising high-CR strategy that also provides a reasonable reconstruction outcome in our cUSi setup. We also see that computational reconstruction can be challenging in both the near and far ends of the imaging range due to near-field artifacts and reduced contrast at greater depths. Therefore, when imaging the carotid artery, a suitable stand-off may help position the vessel at a proper depth for accurate and reliable clinical imaging.

Based on these considerations, a comprehensive in-vivo imaging study will be pursued in the next phase of research, also focusing on CA flow monitoring and Doppler estimation.

## V. Conclusion

In this study, we proposed a 3D cUSi system tailored for the CA imaging, consisting of a 240-element matrix probe with a large $40 \times 24$ mm² footprint and a spatial encoding mask. We detailed the design, characterization, and calibration of the system. Volumetric B-mode images were successfully reconstructed in phantom experiments using Hadamard-encoded MF, with further improved lateral resolution by mask encoding. Furthermore, even with a reduced number of transmissions (16-angle plane waves), the system maintained reasonable image quality, and LSQR-based reconstruction was shown to further enhance performance. These results suggest that the proposed system is suitable for 4D CA imaging.

## Acknowledgment

We thank Stein Beekenkamp and Geert Springeling (Medical Instrumentation, Erasmus MC) for fabricating the encoding mask and fixation frame; Luxi Wei and Robert Beurskens (both at Cardiology, Erasmus MC) for valuable discussions, assistance with data processing and hardware support.

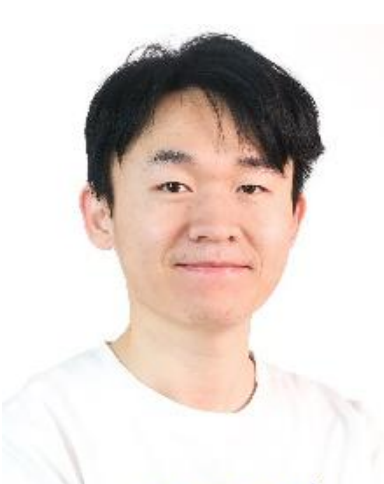

**Yuyang (FOX) Hu** (Student member, IEEE) studied and earned his bachelor's and master's degrees in biomedical engineering from Shenzhen University from 2011 to 2019. He is currently pursuing his Ph.D. degree in the Thoraxcenter Biomedical Engineering group, Department of Cardiology, Erasmus Medical Center, Rotterdam, the Netherlands.

His research is about applying the computational ultrasound imaging for the carotid artery monitoring with low number of sensors.

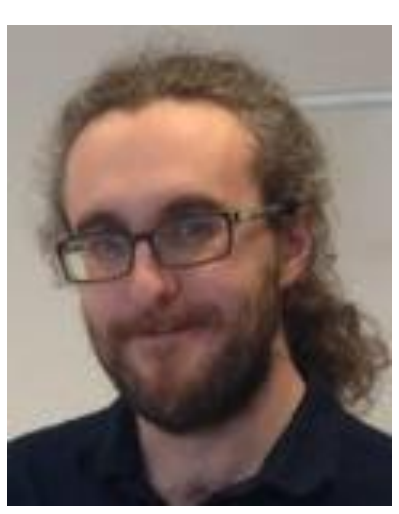

**Michael Brown** (Member, IEEE) got his MSc in physics from the University of Bristol in 2012, MRes in Medical and Biomedical imaging, and PHD from UCL in 2014 and 2018, respectively. He is currently a postdoctoral research fellow in the Department of Medical Physics and Biomedical Engineering at UCL and the Department of Neuroscience at Erasmus MC. His research interests are methods for ultrasound wavefront shaping and computational and functional imaging.

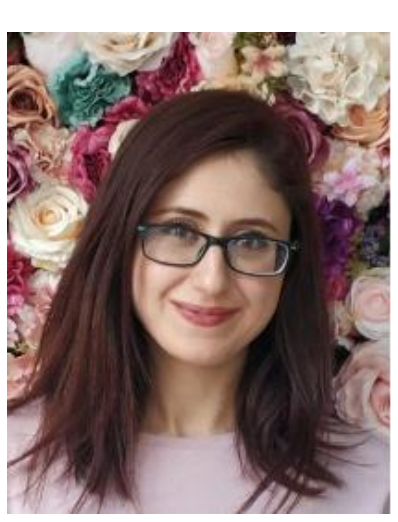

**Didem Dogan** (Member, IEEE) received the M.Sc. degree in electrical engineering in 2020 from Middle East Technical University, Ankara, Turkiye. She is currently pursuing the Ph.D. degree with the Signal Processing Systems department of the Delft university of Technology, Delft, Netherlands. Her research interests include the general area of signal processing, with an emphasis in computational imaging and particularly the ultrasound imaging.

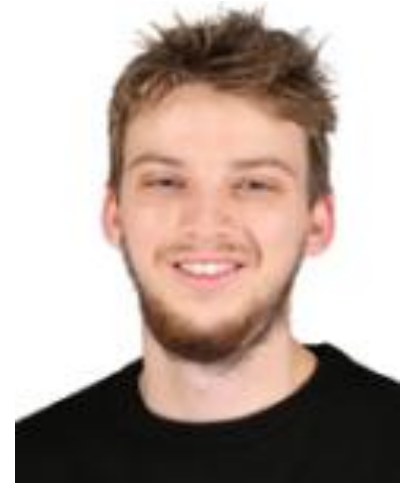

**Mahe Bulot** studied and earned his master's degree in acoustic and vibration engineering in ENSIM school based in Le Mans. He is currently part of Active Probe department in Vermon as acoustic design engineer.

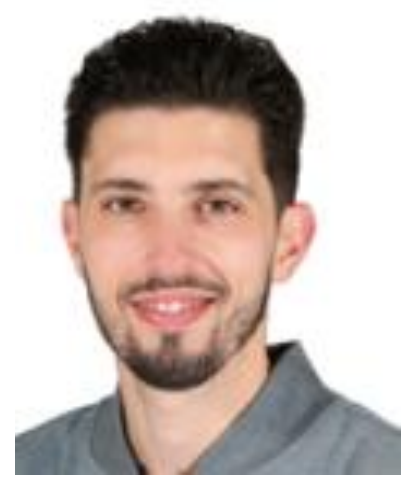

**Maxime Cheppe** obtained his Professional Bachelor's degree in Biomedical Microtechnology Engineering in 2005 from the Tours Polytechnic school. He is a mechanical designer in the Active Probe department at Vermon SA (Tours) since 2013.

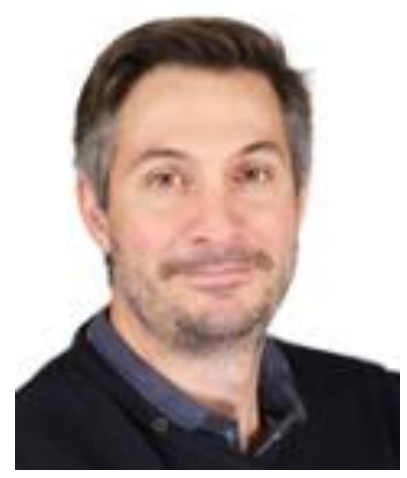

**Guillaume Ferin** is an enthusiastic engineer with a PhD in ultrasound technology for medical imaging from Université François Rabelais (2006). Since 2002, he has been with VERMON in Tours, France, where he focuses on the design and manufacturing of ultrasound transducers. His career includes a period at Vizyontech Imaging (2012-2014) developing advanced 3D ultrasound solutions for early breast cancer detection before returning to VERMON's innovation department. For the last six years, he has led the Active Probes service at VERMON, overseeing the design of innovative 2D and 3D ultrasound devices with embedded electronics.

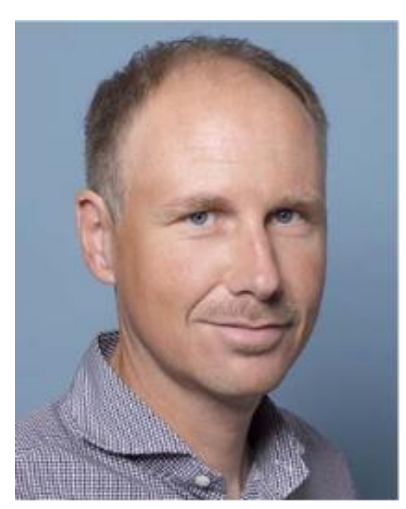

**Geert Leus** (Fellow, IEEE) received the M.Sc. and Ph.D. degrees in electrical engineering from the KU Leuven, Leuven, Belgium, in June 1996 and May 2000, respectively. He is currently a Full Professor with the Faculty of Electrical Engineering, Mathematics and Computer Science, Delft University of Technology, Delft, The Netherlands. He was the recipient the 2021 EURASIP Individual Technical Achievement Award, a 2005 IEEE Signal Processing Society Best Paper Award, and a 2002 IEEE Signal Processing Society Young Author Best Paper Award. He was also Member-at-Large of the Board of Governors of the IEEE Signal Processing Society, Chair of the IEEE Signal Processing for Communications and Networking Technical Committee, Chair of the EURASIP Technical Area Committee on Signal Processing for Multisensor Systems, Editor in Chief of the EURASIP Journal on Advances in Signal Processing, and Editor in Chief of EURASIP Signal Processing. He is a Fellow of EURASIP.

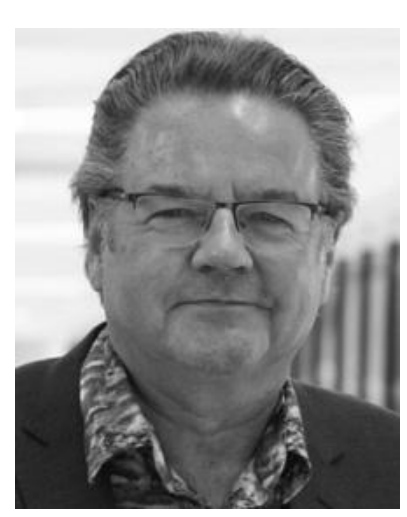

.**Antonius F. W. van der Steen** (Fellow, IEEE) received the M.Sc. degree in applied physics from Technical University Delft, Delft, The Netherlands, in 1989, and the Ph.D. degree in medical science from Catholic University Nijmegen, Nijmegen, The Netherlands, in 1994. He is currently the Head of biomedical engineering with the Thorax Center, Erasmus MC, Rotterdam, The Netherlands. He is an expert in ultrasound, cardiovascular imaging, and cardiovascular biomechanics. He is a fellow of the European Society of Cardiology, member of the Netherlands Academy of Technology (AcTI) and Board Member of the Royal Netherlands Academy of Sciences (KNAW). He was a recipient of the Simon Stevin Master Award and the NWO PIONIER Award in Technical Sciences.

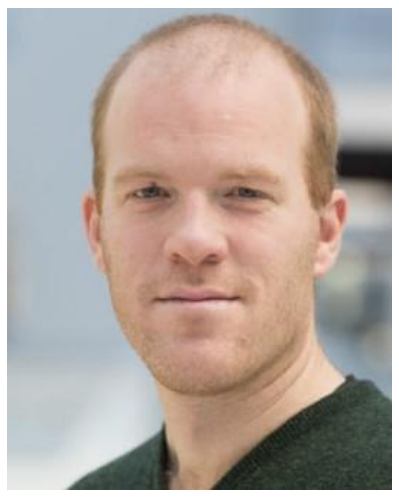

**Pieter Kruizinga** (Member, IEEE) received the Ph.D. degree from Erasmus MC, Rotterdam, The Netherlands, in 2015.

In 2018, he joined the Neuroscience Department, Erasmus MC, where he leads the imaging research with the Center for Ultrasound and Brain-Imaging Erasmus MC (CUBE). His current research focuses on computational ultrasound imaging and functional ultrasound imaging of the brain.

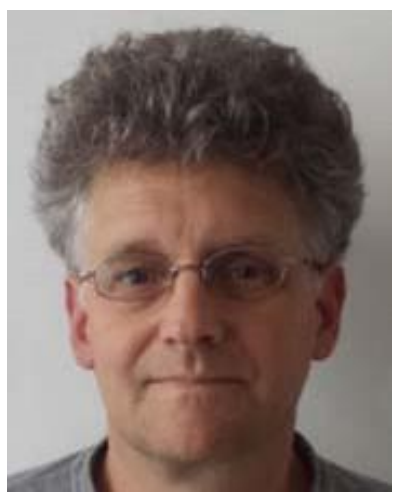

**Johan G. Bosch** (Member, IEEE) received the M.Sc. degree in electrical engineering from the Eindhoven University of Technology, Eindhoven, The Netherlands, in 1985, and the Ph.D. degree from the Leiden University Medical Center, Leiden, The Netherlands, in 2006.

He is currently an Associate Professor and a Staff Member with Thoraxcenter Biomedical Engineering, Department of Cardiology, Erasmus University Medical Center, Rotterdam, The Netherlands. His research interests include 2D and 3D echocardiographic image formation and processing, transducer development, and novel ultrasound techniques for image formation and functional imaging.

# Supplementary Material

# A 240 Elements Matrix Probe with Aberration Mask for 4D Carotid Artery Computational Ultrasound Imaging

Yuyang Hu[1]*, Michael Brown[2], Didem Dogan[3], Mahé Bulot[4], Maxime Cheppe[4], Guillaume Ferin[4], Geert Leus[3], Antonius F.W. van der Steen[1], Pieter Kruizinga[2, 3, 5], Johannes G. Bosch[1]

[1] Department of Cardiology, Erasmus MC University Medical Center, 3000 CA Rotterdam, The Netherlands.

[2] Department of Neuroscience, Erasmus MC University Medical Center, 3000 CA Rotterdam, The Netherlands.

[3] Department of Micro Electronics, Delft University of Technology, 2628 CJ Delft, The Netherlands.

[4] Active Probe Group, Innovation Department, Vermon SA, Tours, France.

[5] Department of Neurosurgery, Erasmus MC University Medical Center, 3000 CA Rotterdam, The Netherlands.

*Corresponding author, E-mail address: y.hu@erasmusmc.nl (Y. Hu).

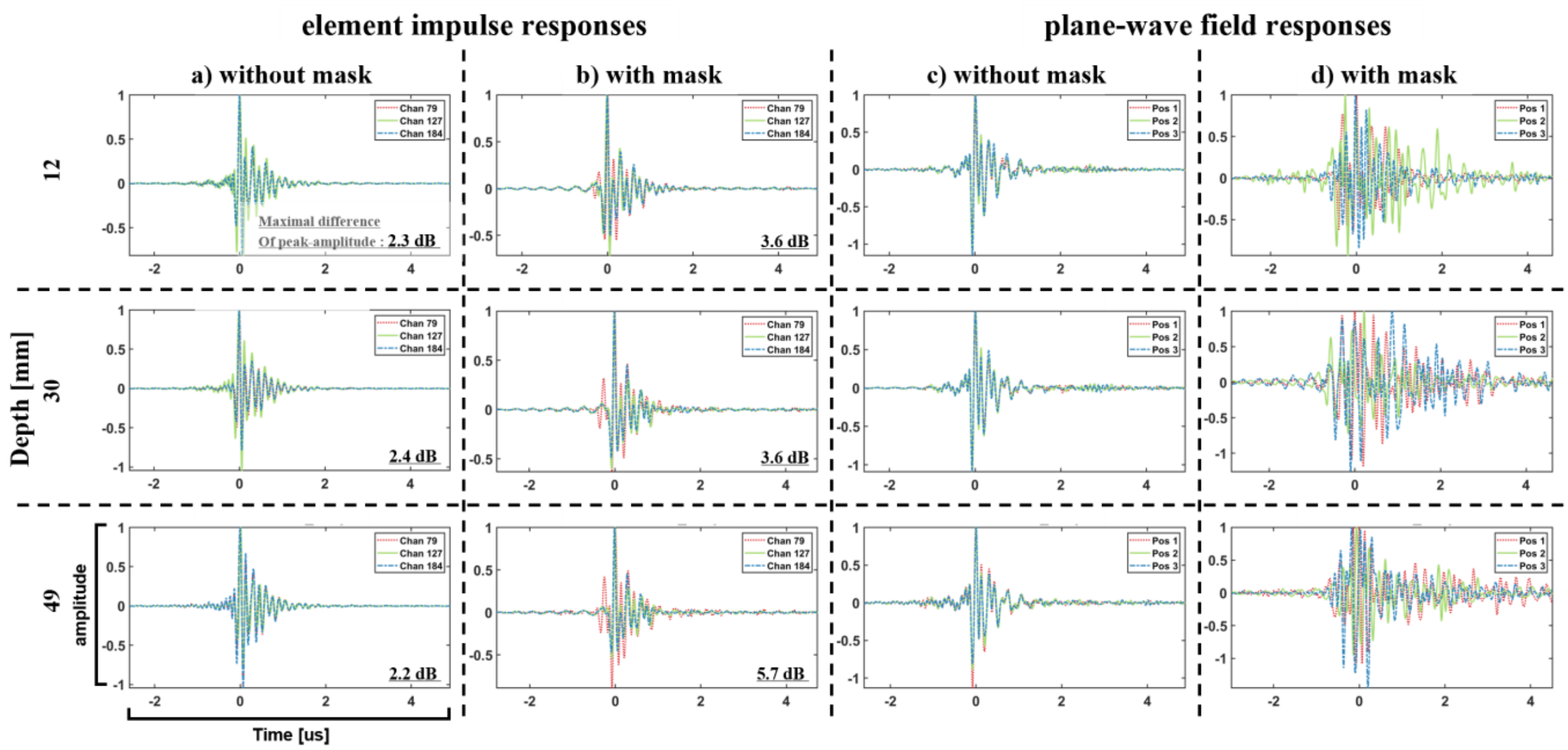

Fig. S1 Comparison of element-wise impulse responses and forward plane-wave field responses with and without the mask. a–b) Impulse responses of three randomly selected elements evaluated at multiple depths, illustrating the element-to-element variations introduced by the mask. c–d) Forward plane-wave field responses sampled at the same three lateral positions across different depths, showing that the mask induces strong depth- and position-dependent variations, whereas the no-mask case remains highly similar. All responses are normalized and aligned to their peak amplitude here for visual shape comparison. Maximal inter-element differences in peak amplitude are given in the inset in dB (bottom right per subfigure in a-b). The inter-element amplitude differences with the mask (max 5.7 dB at 49mm) were only slightly larger than without mask (lower than 2.4 dB in all depths).

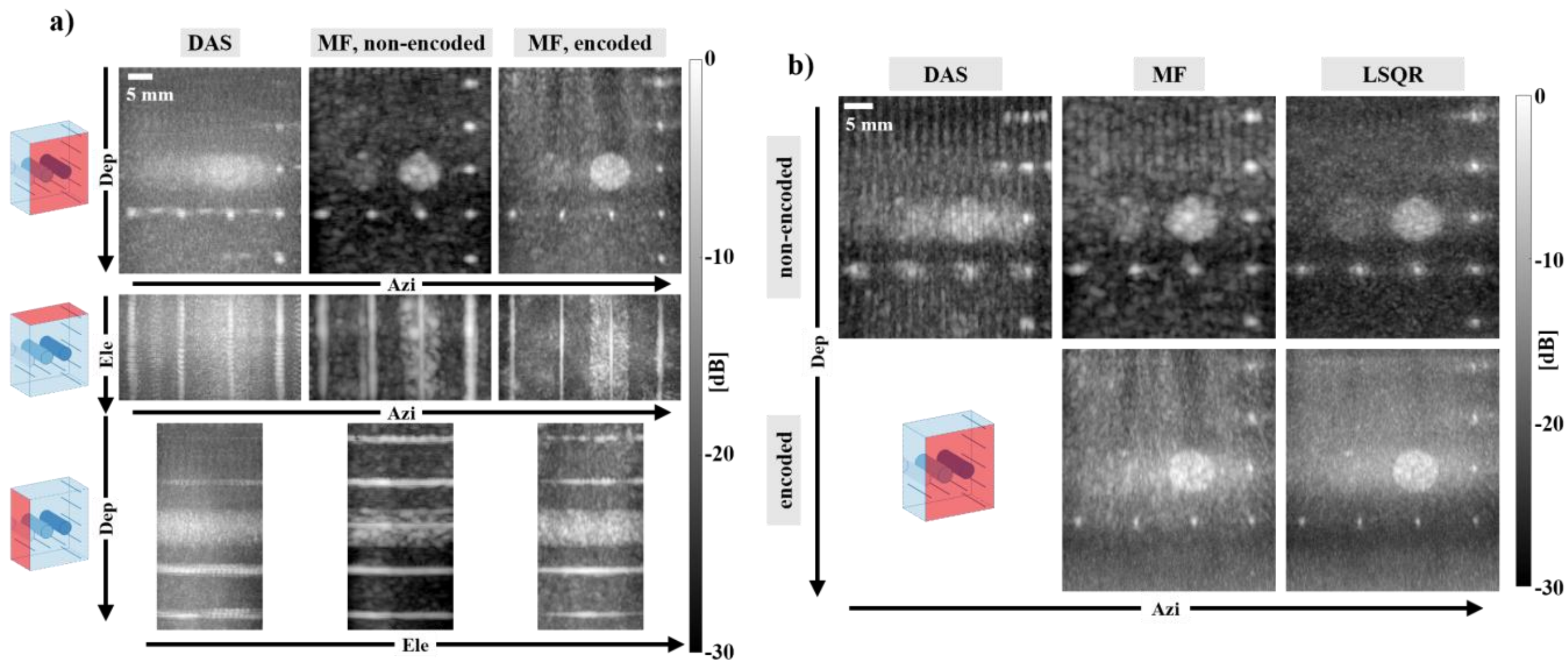

Fig. S2 Volumetric reconstructions without boundaries of inclusion 1 and inclusion 2 marked. a) Hadamard transmit scheme, corresponding to Fig. 5a; b) 16-angle plane-wave transmit scheme, corresponding to Fig. 6a.

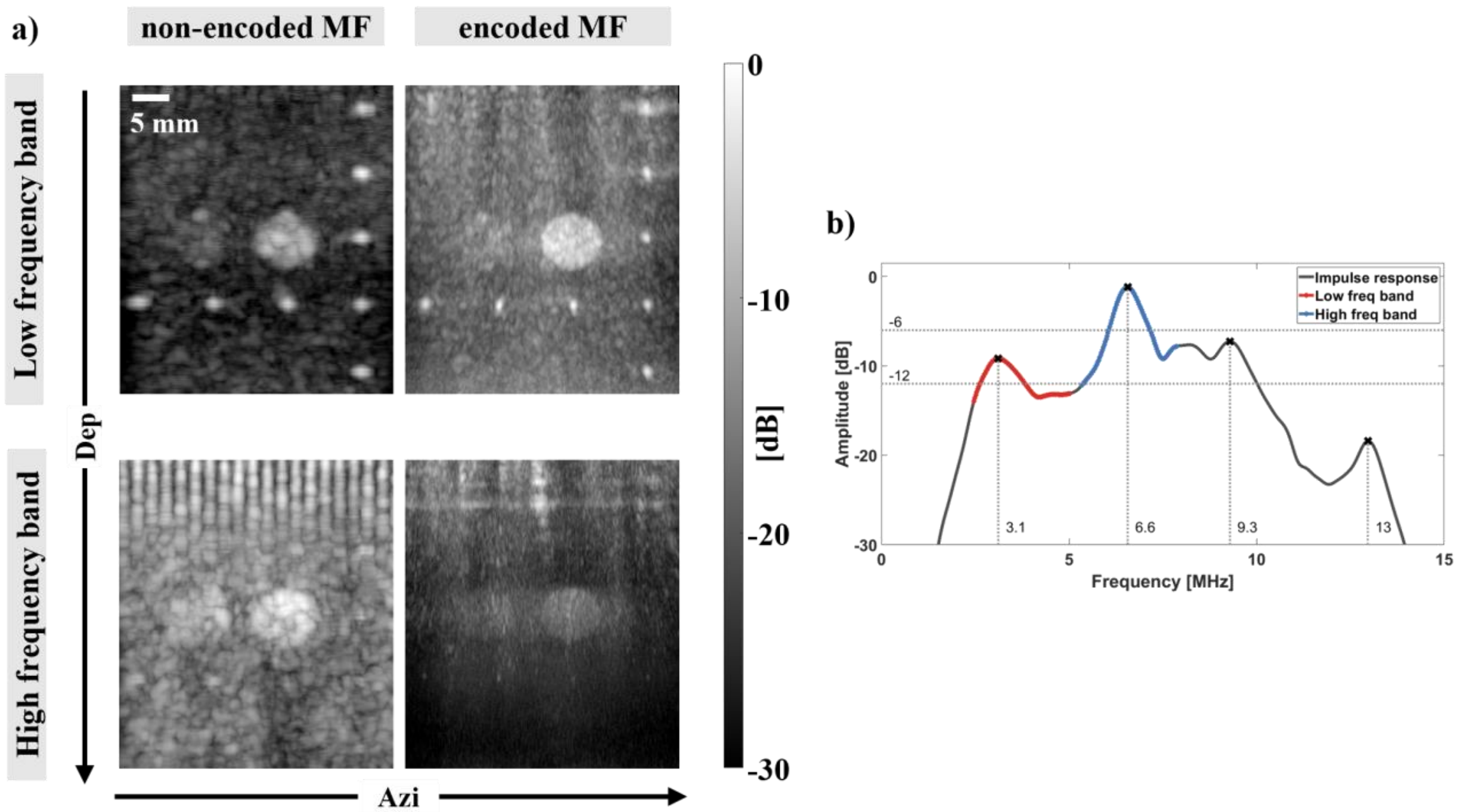

Fig. S3 a) Comparison of reconstructions using different frequency bands, under the Hadamard transmit scheme with MF reconstruction. Volumetric images are shown as MIPs along the elevation domain. The top row shows reconstruction using the low-frequency band (2.4-5.0 MHz); the bottom row shows results using a higher-frequency band (5.3-7.9 MHz) around the probe's center frequency. Reconstruction quality degrades in the higher-frequency band. Without encoding (left column), wire targets become difficult to distinguish, likely due to severe spatial undersampling. With encoding (right column), wire targets appear sharper but show significantly lower contrast, possibly due to greater delay and attenuation of high-frequency components in the mask, leading to sharper PSFs but elevated background levels. b) Frequency spectra indicating the selected low (red) and high (blue) bands.

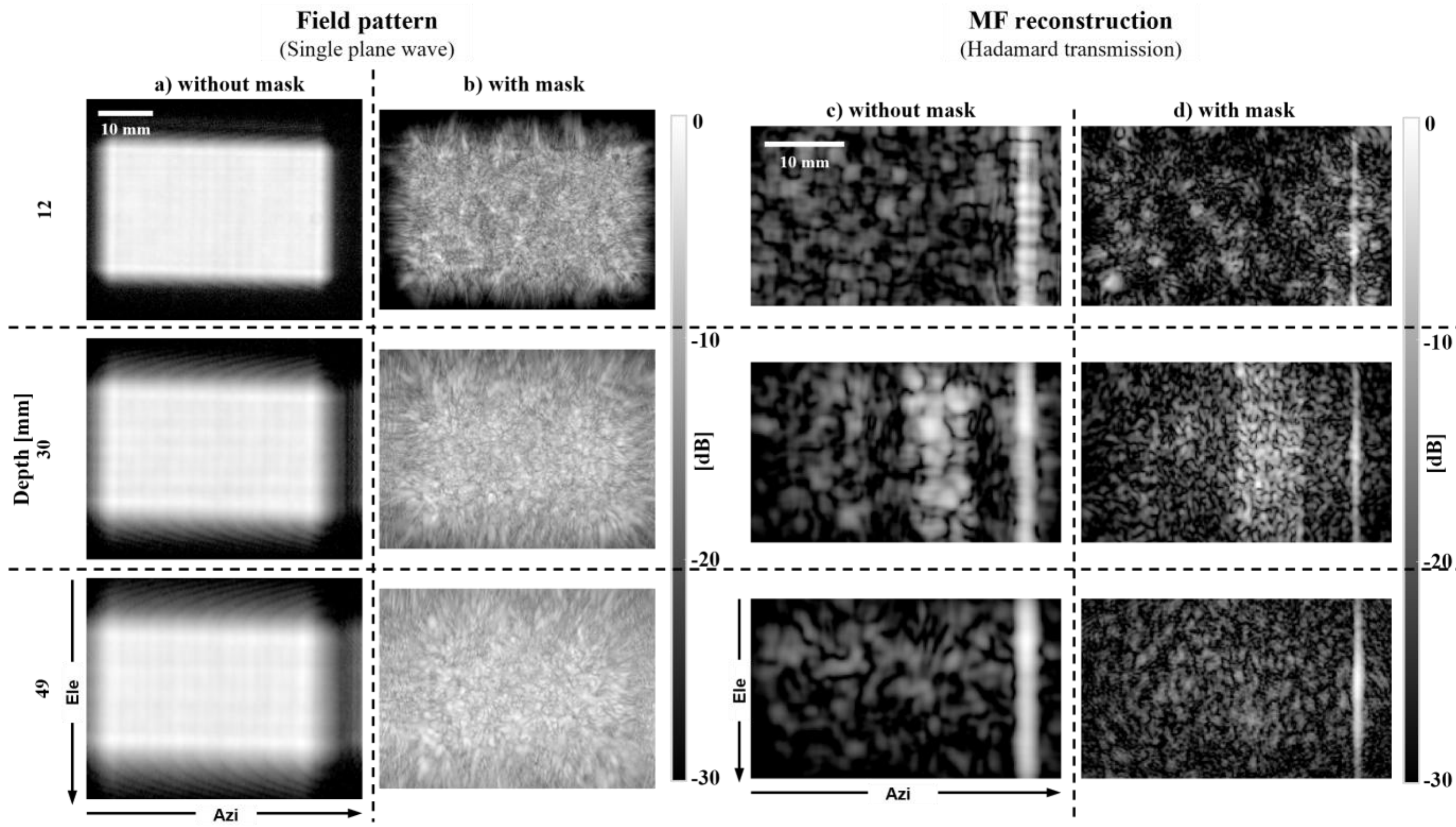

Fig. S4 a–b) Forward field patterns generated by a single plane-wave transmission at 12 mm, 30 mm, and 49 mm depths. These figures can be compared with Fig. 4c and 4f. Without the mask (left column), the fields remain smooth and highly similar across depths; with the mask (right column), the fields are encoded and become more spread out (overlapping) with increasing depth, while also showing decreased contrast. c–d) MF reconstructions of the CIRS phantom at the same depths using Hadamard transmissions. These figures can be compared with Fig. 5a. Wire targets can be distinguished and visually compared. With-mask provides significantly sharper reconstructions. Deeper reconstructions appear smoother than at shallow depth both without-mask and with-mask.

**Other Supplementary Materials: Movies:**

Movie M1: Comparison of different 3D reconstructions of CIRS phantom after Hadamard transmission. Columns from left to right: DAS, non-encoded MF, and encoded MF reconstructions.

Movie M2: Comparison of different 3D reconstructions of CIRS phantom after 16-angle plane wave transmission. Columns from left to right: DAS, MF, and LSQR reconstructions. The top row shows cases without encoding, while the bottom row shows cases with encoding (mask).